\documentclass{PoS}

\usepackage[caption=false]{subfig}
\usepackage[sort&compress,numbers]{natbib}
\usepackage[T1]{fontenc}
\usepackage{lmodern}
\usepackage{soul}
\usepackage{enumitem}
\usepackage{xcolor}
\usepackage[titletoc]{appendix}
\usepackage{overpic}
\usepackage[normalem]{ulem}
\usepackage[nointegrals]{wasysym}

\usepackage[caption=false]{subfig}
\usepackage{graphicx}
\usepackage{nicefrac}
\usepackage{tabularx}
\usepackage{booktabs}
\usepackage{multirow}
\usepackage{amssymb,amsmath}
\usepackage{footnote}
\usepackage{bm}
\usepackage[capitalize]{cleveref}[2012/02/15]

\usepackage{accents}

\usepackage{lineno}
\usepackage{tabu}
\usepackage{url}
\usepackage{siunitx}
\DeclareSIUnit\parsec{pc}
\DeclareSIUnit\lightyear{ly}

\let\OldAng\ang%
\renewcommand*{\ang}[2][]{%
    \OldAng[scientific-notation=false,#1]{#2}%
}
\DeclareSIUnit\year{yr}
\DeclareSIUnit\erg{erg}
\DeclareSIUnit\msun{M_{\astrosun}}
\DeclareSIUnit{\GeV}{\giga\electronvolt}
\DeclareSIUnit{\TeV}{\tera\electronvolt}
\DeclareSIUnit{\PeV}{\peta\electronvolt}
\DeclareSIUnit{\MeV}{\mega\electronvolt}
\DeclareSIUnit{\eV}{\electronvolt}
\DeclareSIUnit{\smm}{\square\metre\second}
\DeclareSIUnit{\smmr}{\metre^{-2}\second^{-1}}
\DeclareSIUnit{\dc}{d.c.}
\DeclareSIUnit{\pe}{p.e.}
\DeclareSIUnit{\nucleon}{nucleon}

\definecolor{desyOrange}{RGB}{242,142,0}
\usepackage{csquotes}% Recommended
\title{Joint Likelihood Fits for the Study of Galactic Objects with HAWC}

\ShortTitle{HAWC Fits}

\author{\speaker{Henrike Fleischhack}{}\\
        Michigan Technological University%, 1400 Townsend Drive, Houghton, MI, 49931, USA
        \\
        E-mail: \email{hfleisch@mtu.edu}}

\author{Petra Huentemeyer\\
        Michigan Technological University
}
\author{for the HAWC collaboration \thanks{See \protect\url{http://www.hawc-observatory.org/collaboration/icrc2017.php} for full author list.}}

\abstract{Studying gamma-ray emission by Galactic objects is key to understanding the origins and acceleration mechanisms of Galactic cosmic ray electrons and hadrons. The HAWC observatory provides an unprecedented view of the gamma-ray sky at TeV energies and is particularly suited for the study of Galactic objects. However, the interpretation of the measured data poses several challenges. The high density of sources and source candidates can cause source confusion and make it harder to disentangle the origin of the emission. The relatively low angular resolution of HAWC, compared to instruments in optical or radio wavelengths, can further cause the emission of neighboring sources to bleed into each other or even make them look like one extended source. On the other hand, with its wide field of view, HAWC is uniquely suited for the study of extended sources. However, this requires the simultaneous modeling of both their morphology and emission spectrum. Joint likelihood fits to data taken over a larger range of energies can help overcome these challenges and achieve the full potential of the HAWC detector. In this presentation, we will discuss how systematic uncertainties related to joint likelihood fits can affect the measurements. }

\FullConference{35th International Cosmic Ray Conference --- ICRC2017\\
		10--20 July, 2017\\
		Bexco, Busan, Korea}

\begin{document}

\section{Introduction}
The HAWC detector has been observing the very-high energy (VHE, $E>\SI{100}{\GeV}$) sky for more than two years. The first catalog, comprising 39 sources, was released in early 2017 \cite{2HWC}. With a large dataset collected, HAWC is starting to become systematics-limited for some studies. In this paper, we will investigate two sources of systematic uncertainty: the correlation between assumed source location and flux measurements as well as the uncertainty in flux measurements due to incomplete modeling of the point spread function (PSF).

\section{The HAWC Detector}
The High Altitude Water Cherenkov Observatory (HAWC) \cite{2HWC} is a wide-field-of-view gamma-ray detector located at an altitude of \SI{4.1}{\kilo\meter} in the volc\'an Sierra Negra, Mexico. It is sensitive to gamma-ray showers above \SI{300}{\GeV} and is able to observe $\gamma$-ray sources at declinations between \ang{-20} and +\ang{60}, with an instaneous field-of-view of nearly \SI{2}{\steradian}. 

HAWC consists of 300 water cherenkov detectors (WCDs), each instrumented with four photomultiplier tubes (PMTs). Charged particles in extensive air showers emit Cherenkov light in the WCDs, which is detected by the PMTs. The readout system uses discriminators with two different thresholds, with the lower (higher) threshold corresponding to a pulse height of about 0.25 (4.0) photo-electrons. When a readout is triggered, usually requiring 28 PMTs to be hit within a time window of \SI{150}{\nano\second}, the timestamps of the threshold crossings for each PMT are written to disk. %Different trigger conditions were used for running with the partial array during reconstruction.

With the full detector online and current trigger settings, the array trigger rate is about \SI{20}{\kilo\hertz}. The event rate is dominated by cosmic-ray showers by several orders of magnitude, even for strong gamma-ray sources. Most of this background can be removed later during data analysis using cuts on the shape of the lateral distribution function of the shower.

\section{Data Analysis}
\subsection{Event Reconstruction}
The event reconstruction proceeds in several steps. First, the deposited charge and arrival time in each PMT are calculated from the threshold crossing times.

The shower core position and the shower direction are calculated in an iterative procedure. The shower core is found by fitting the measured charge distribution with a simplified NKG function. The first iteration of the fit uses a simple estimate of the charge-weighted center as a starting value for the core position.

The arrival direction is found by fitting a plane to the shower arrival times in each PMT. The arrival times are corrected for the shower curvature, which causes particles far from the shower core to be delayed due to a longer traveling distance, and for the fact that PMTs that see a high rate tend to measure an earlier arrival time than PMTs that see a low rate of particles (sampling bias). 

The core and direction fits are repeated, using the results from the previous fit as starting values and restricting the PMTs used to hits recorded within \SI{50}{\nano\second} of the previously fitted shower plane.

HAWC currently uses two photon/hadron separation parameters, which are both related to the shape of the lateral distribution function.

The HAWC analysis currently uses the fraction $f_{hit}$ of hit PMTs relative to the number of available, i.e. installed and working PMTs as an energy proxy. As the detector response (e.g., point spread function and detector efficiency) depends on $f_{hit}$, events are binned according to $f_{hit}$ for the following analysis steps. Photon/hadron cuts are optimized separately for each bin to yield the maximum significance on the Crab Nebula. More details on the event reconstruction, $f_{hit}$ bins, and photon/hadron cuts can be found in \cite{HAWCCrab}.

For the results presented here, we used 706 days of HAWC data. The recorded events were binned according to $f_{hit}$ and filled into maps using the HealPix pixelation scheme with $N_{side}=1024$.

\subsection{Likelihood Analysis}
\label{sec:analysis}
The 3ML framework\footnote{\protect\url{https://github.com/giacomov/3ML}}\citep{3ml} was used for data analysis of the HAWC data. It relies on the HAWC likelihood analysis framework \emph{LiFF} \cite{liff}. For the likelihood analysis, the predicted counts from a source plus background model in each bin, accounting for the detector response, are compared to the observed events. The parameters of the source model are adjusted to maximize the likelihood function.

In order to probe the statistical significance of a source candidate, the likelihood of the best-fit source plus background model, $\mathcal{L}^{max}$, is compared to the background-only hypothesis, $\mathcal{L}^{BG}$. We use a test statistic defined as 

\begin{align*}
TS=2 \cdot \log\left( \frac{\mathcal{L}^{max}}{\mathcal{L}^{BG}} \right).
\end{align*} 

In case of nested models with one free parameter the square root of the test statistics corresponds to a gaussian significance. This is the case for the skymaps shown here, where for each pixel one point source was fit on top of the background with the normalization free, but fixed spectral shape.

The source model generally consists of a spatial model and a spectral model. The models are convolved with the appropriate detector response functions and the exposure time to calculate the predicted number of signal events per bin. The number of background counts in each bin is estimated using the \emph{direct integration} method \cite{DI}.

%For the analysis of the Crab nebula presented here, a point source assumption was used for the spatial model and the spectrum was modeled as a log-parabola spectrum,

%\begin{align*}
%\frac{\mathrm{d}N}{\mathrm{d} E \mathrm{d} t \mathrm{d} A } = K\cdot\left(\frac{E}{E_0}\right)^{-\alpha-\beta\cdot \log\left( \frac{E}{E_0} \right)},
%\end{align*}
%with pivot energy $E_0=\SI{7}{\TeV}$ following \cite{HAWCCrab}.

Skymaps (flux maps or significance maps) are created via a putative source search: For each pixel in the map, a single point-source fit is performed with the location of the point source centered on the pixel in question. The spectrum is fixed to a power law  with a pivot energy of \SI{7}{\TeV} and an index of \num{-2.63} (Crab-like). The only free parameter is the flux normalization. The flux map is produced from the best-fit flux normalization values. The significance value for each pixel is calculated from the test statistics comparing the best-fit model to the background-only hypothesis.

\subsection{Detector Resolution}
The likelihood analysis described above requires a model of the detector response, in order to calculate the number of expected counts in each spatial and $f_{hit}$ bin. In particular, it is necessary to know HAWC's point spread function (PSF) to predict how far from the source a gamma-ray event may be reconstructed. HAWC's PSF depends on the fraction of PMTs hit as well as on the source declination. It is determined from simulations of gamma-ray showers, which are reconstructed in the same way as data, and can be described by a double Gaussian function. 

HAWC's angular resolution improves dramatically with the fraction of PMTs hit. For the Crab Nebula and other sources culminating close to zenith, the angular resolution (\SI{68}{\percent} containment radius of the PSF) is about \ang{1} in the first $f_{hit}$ bin (between \SIrange{6.7}{10.5}{\percent} of PMTs hit) and drops to \ang{0.17} in the highest $f_{hit}$ bin (more than \SI{84}{\percent}) of PMTs hit) \cite{HAWCCrab}. The predicted angular resolution from simulations was compared to the width of the gamma-ray emission from the Crab Nebula, whose true extent is assumed to be much smaller than HAWC's PSF. The two agree within the statistical uncertainties on the measured PSF, which correspond to \SIrange{5}{10}{\percent} \cite{HAWCCrab}.

The absolute pointint of the HAWC detector was calibrated with the known position of the Crab Nebula, and cross-checked using the known positions of two other bright point sources, the blazars Mrk 421 and Mrk 501. The residual uncertainty in the absolute pointint was estimated to be about \ang{0.1} \cite{HAWCCrab}.

The standard HAWC analysis currently uses the fraction of PMTs hit as an energy proxy. The resolution of this energy proxy is very poor and there is large overlap in the energy distributions of the different $f_{hit}$ bins. Still, there is sufficient correlation between $f_{hit}$ and energy that HAWC is sensitive to the spectral shape via the $f_{hit}$ dependent likelihood analysis. Energy estimators with improved resolution have been developed and are currently being evaluated \cite{HAWCenergy}. 

\subsection{Source Injection Studies}
In order to study the effect of an inadequately modeled PSF on the analysis results, the width of the HAWC PSF model was scaled up and down within the uncertainties mentioned above. Nine `scaled' PSF models were produced, with scaling factors between 0.8 and 1.2. The nominal HAWC PSF is modeled as the sum of two Gaussians for each $f_{hit}$ bin. For the scaled models, both Gaussian width parameters were multiplied by the same scaling factor. 

For each scaled PSF model, ``fake'' maps were produced by injecting point sources on top of the background (from direct integration) and applying Poissonian fluctuations. The spectrum of the injected source was fixed to the measured Crab spectrum from \cite{HAWCCrab}, with a log-parabola shape:

\begin{align*}
\frac{\mathrm{d}N}{\mathrm{d}E} = N_{0}\cdot \left( \frac{E}{E_0} \right)^{-\alpha+\beta\cdot\log\left(E/E_0\right)},
\end{align*}
with pivot energy $E_0=\SI{7}{\TeV}$, flux normalization $N_0=\SI[exponent-to-prefix=false,scientific-notation=true]{2.512e-13}{\per\TeV\per\second\per\centi\metre\squared}$, spectral index $\alpha=2.63$ and curvature parameter $\beta=-0.15$.

Different offsets from the nominal position of the Crab Nebula were tried.  For each value of the PSF scaling factor and each value of the source offset, twenty maps were produced with different random seeds for the fluctuations. A Crab-like source injected with the nominal PSF is detected with $\sqrt{TS}>130$ compared to the background-only model.

\section{Correlation Between Source Offset, PSF, and Measured Spectrum}
\begin{figure}[ptb]
\subfloat[]
	[Dependence of the flux normalization on the offset of the injected source.]
	{\includegraphics[width=0.5\textwidth]{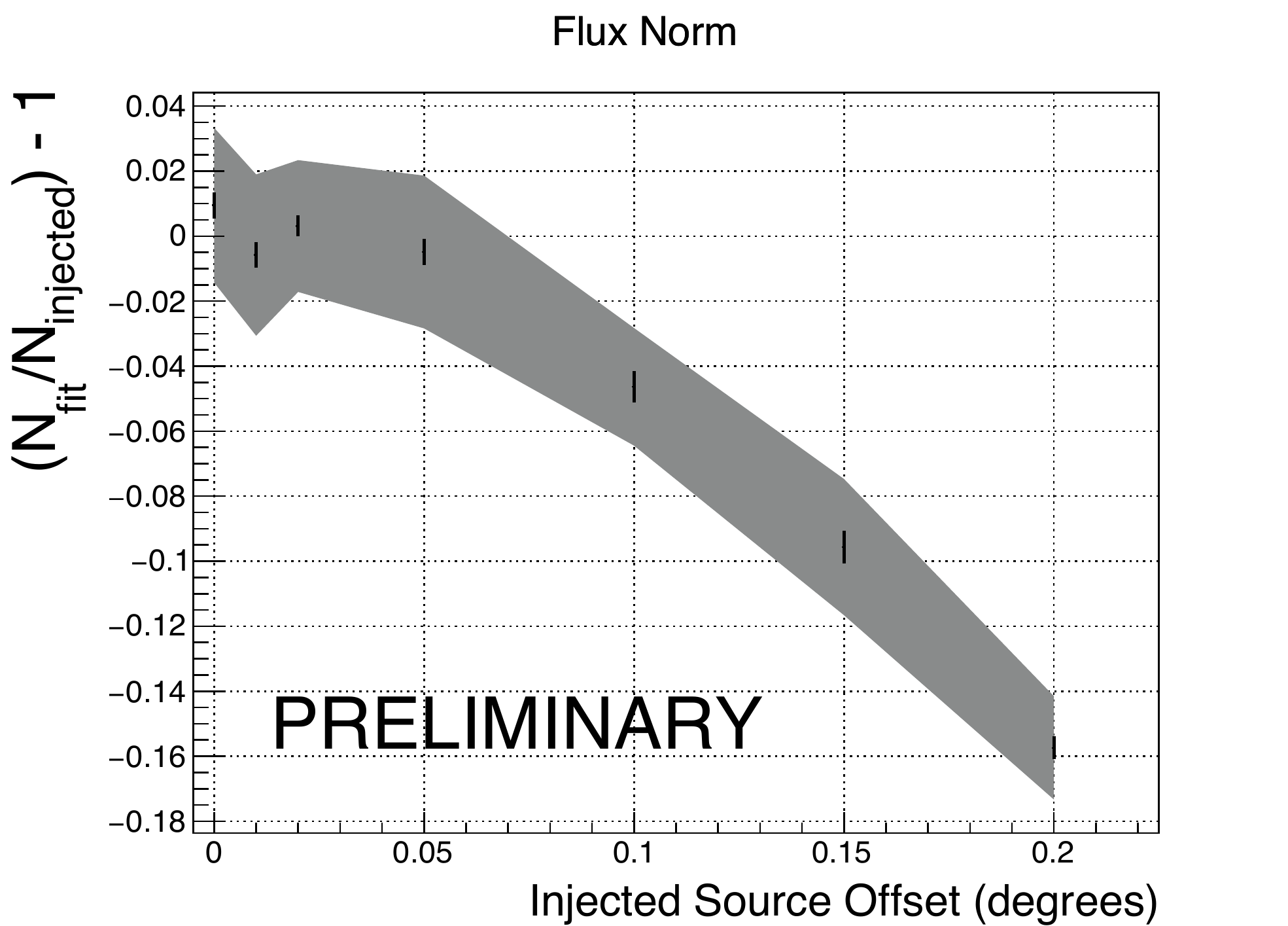}}
\subfloat[]
	[Dependence of the spectral index on the offset of the injected source]
	{\includegraphics[width=0.5\textwidth]{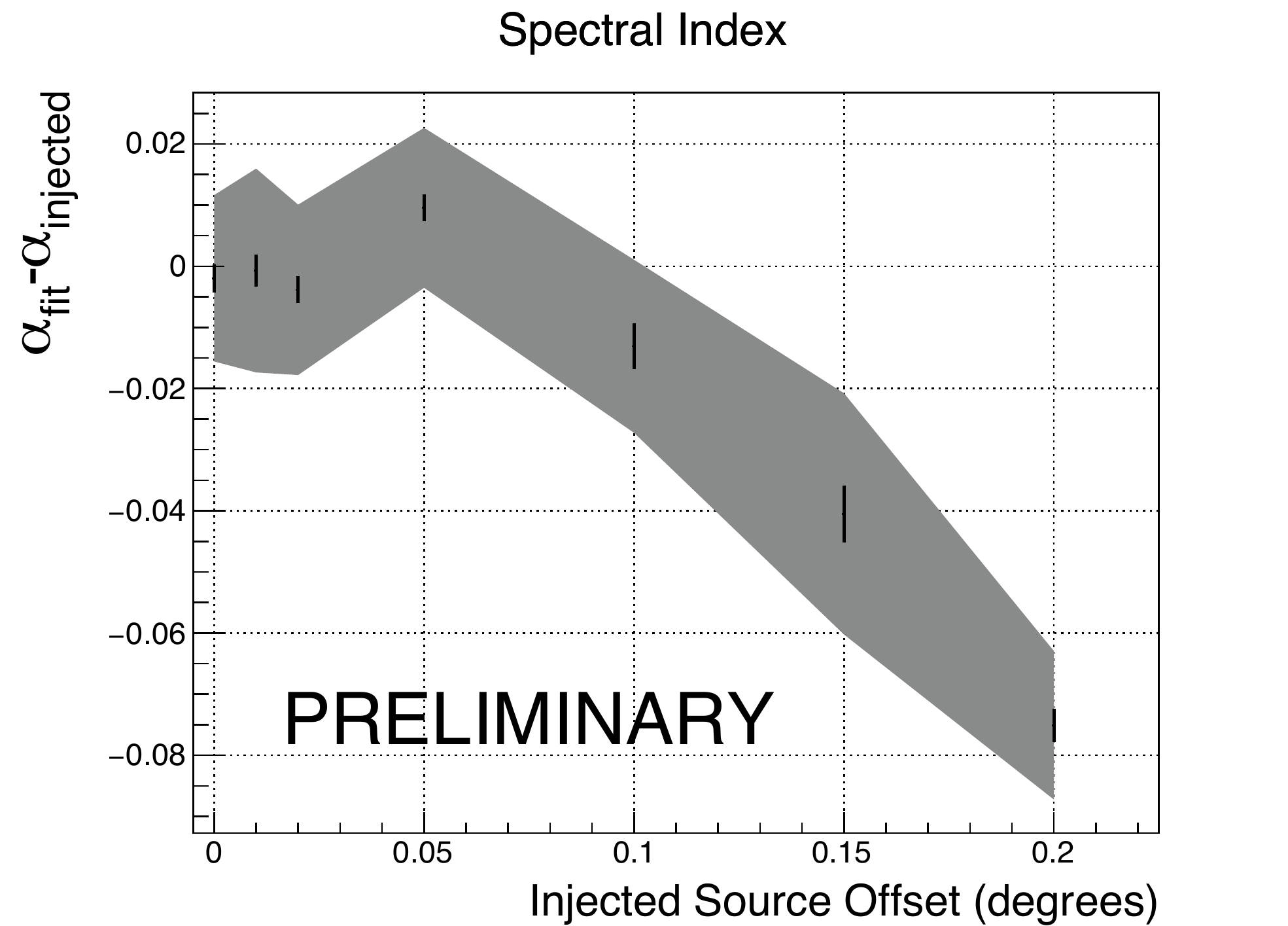}}
\\
\subfloat[]
	[Dependence of the flux normalization on the width of the injected source.]
	{\includegraphics[width=0.5\textwidth]{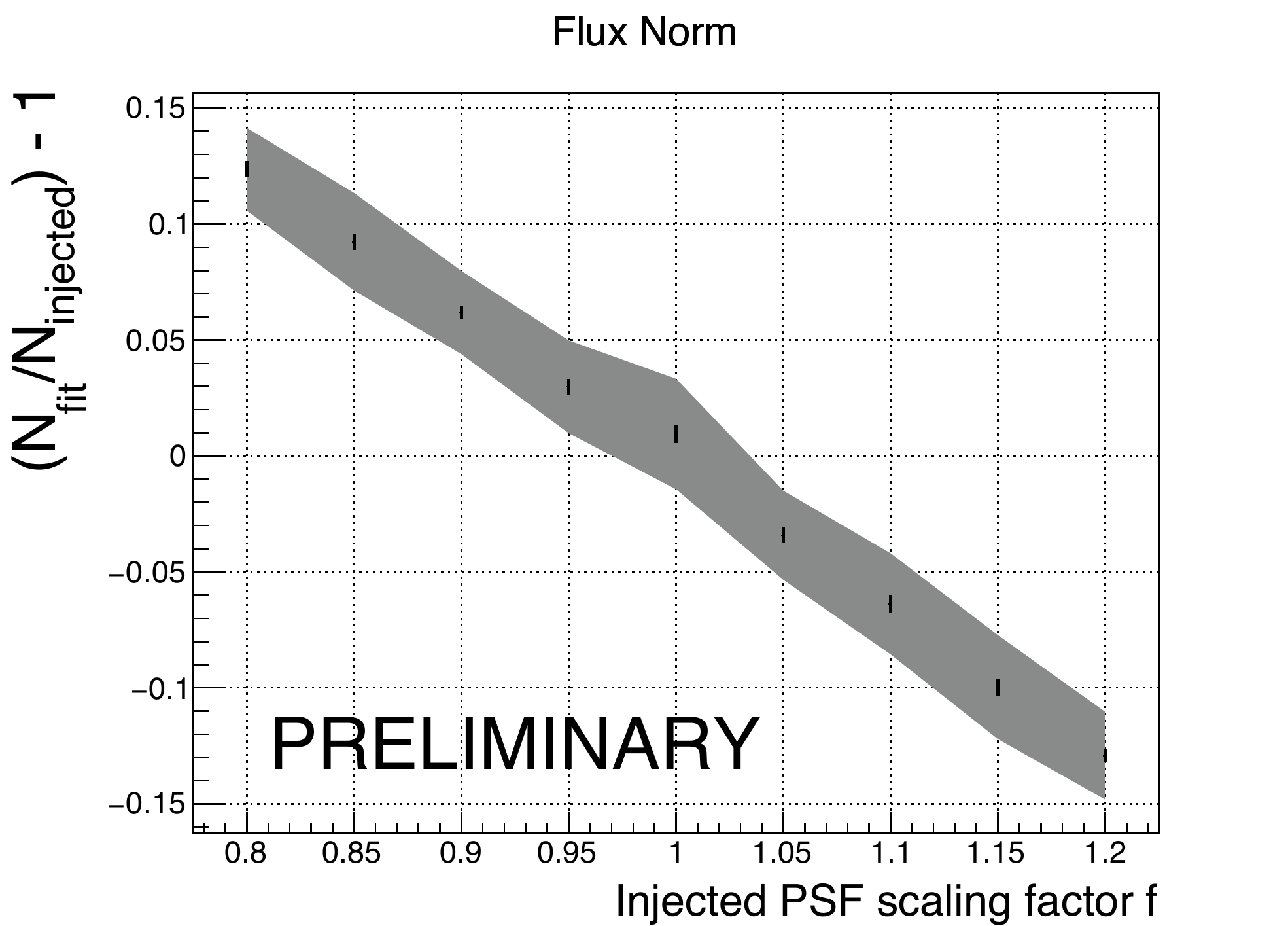}}
\subfloat[]
	[Dependence of the spectral index on the width of the injected source]
	{\includegraphics[width=0.5\textwidth]{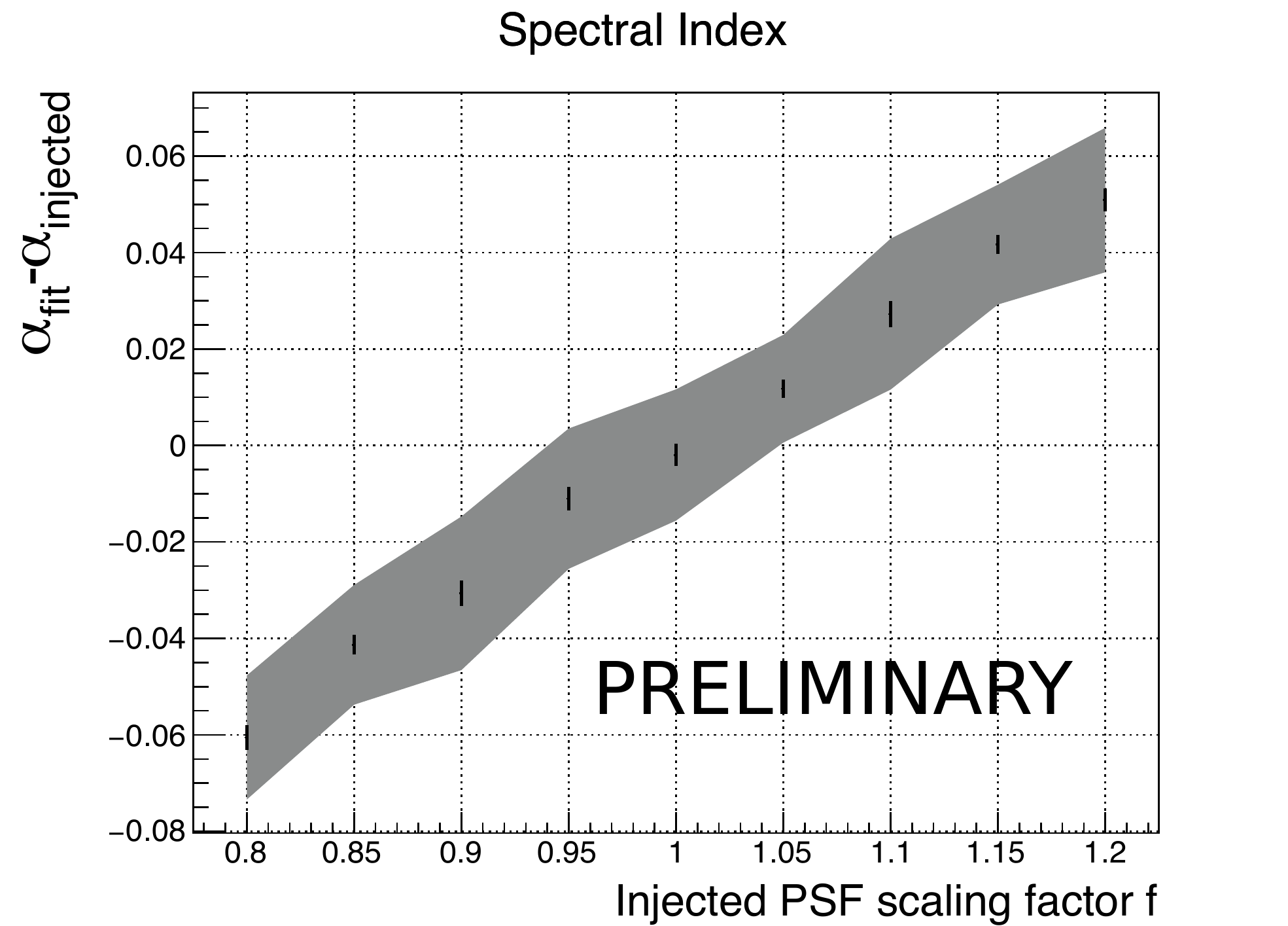}}
\caption{Spectral parameters depending on the scaling factor applied to the PSF and the offset of the injected source from the nominal Crab position. Black bars: Statistical uncertainty on the mean, grey band: width of the distribution.}
\label{fig:results}
\end{figure}

The injected source maps produced according to the prescription above were fit with a point source model at the nominal position of the Crab Nebula and a log-parabola spectrum. The spectral curvature parameter $\beta$ did not show any significant dependence on either the source offset or the injected witdh. The resulting distributions of the best-fit flux normalization and spectral index can be seen in \cref{fig:results}. Within the statistical uncertainties, the spectral fit does not change significantly if the source is offset by less than \ang{0.1} from the nominal position. However, there is a dependence on the width of the injected source: For an injected PSF that is \SI{10}{\percent} narrower than the PSF used for the analysis, the recontructed spectrum is (on average) softer by 0.03 and has a \SI{6}{\percent} higher flux normalization. If the injected PSF is \SI{10}{\percent} wider than nominal, the flux is underestimated and the index becomes harder. These effects are already accounted for in the systematic uncertainties. HAWC assesses a \SI{20}{\percent} uncertainty on the measured flux normalization and an uncertainty of $0.1$ on the spectral index due to incomplete modeling of the point spread function.

\section{Spatial Residuals}
\begin{figure}[p]
\subfloat[]
	[PSF scaled by 0.85.]
	{\begin{overpic}[width=0.36\textwidth]{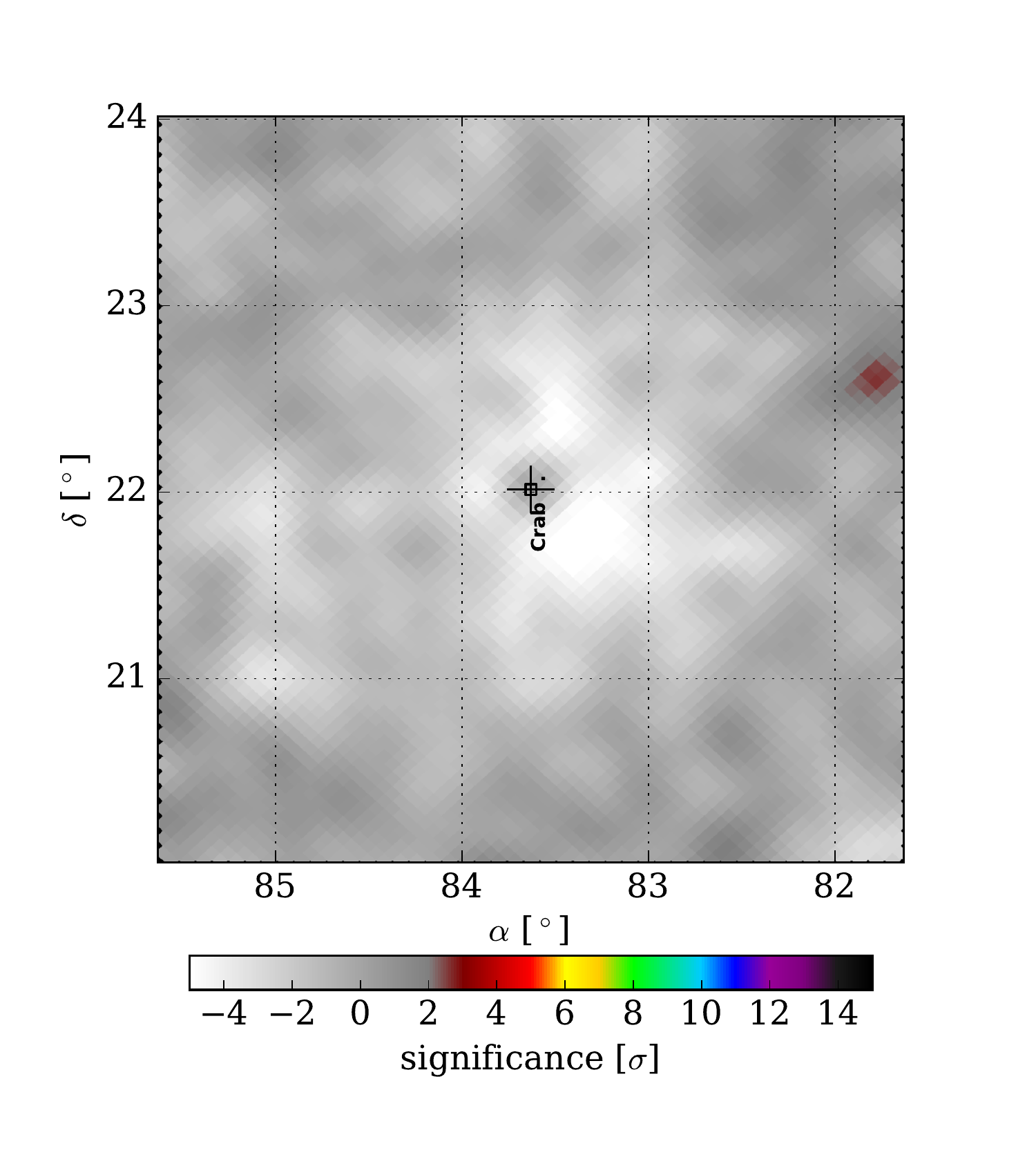}\put(15,83){\color{magenta}{\textbf{PRELIMINARY}}}\end{overpic}\label{fig:f09}}
\subfloat[]
	[PSF scaled by 1.15.]
	{\begin{overpic}[width=0.36\textwidth]{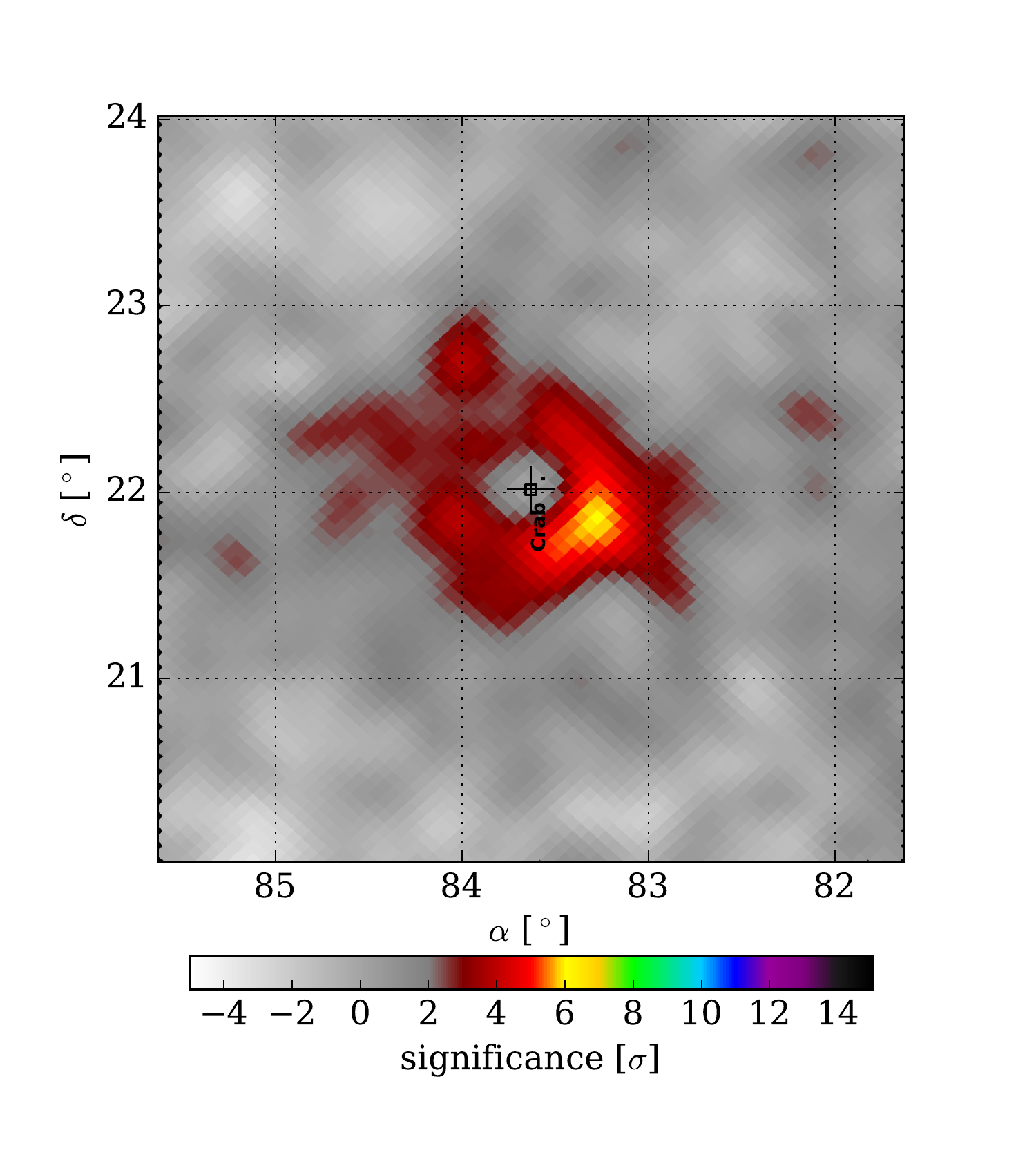}\put(15,83){\color{magenta}{\textbf{PRELIMINARY}}}\end{overpic}\label{fig:f11}}

\subfloat[]
	[Source offset by \ang{0.1}.]
	{\begin{overpic}[width=0.36\textwidth]{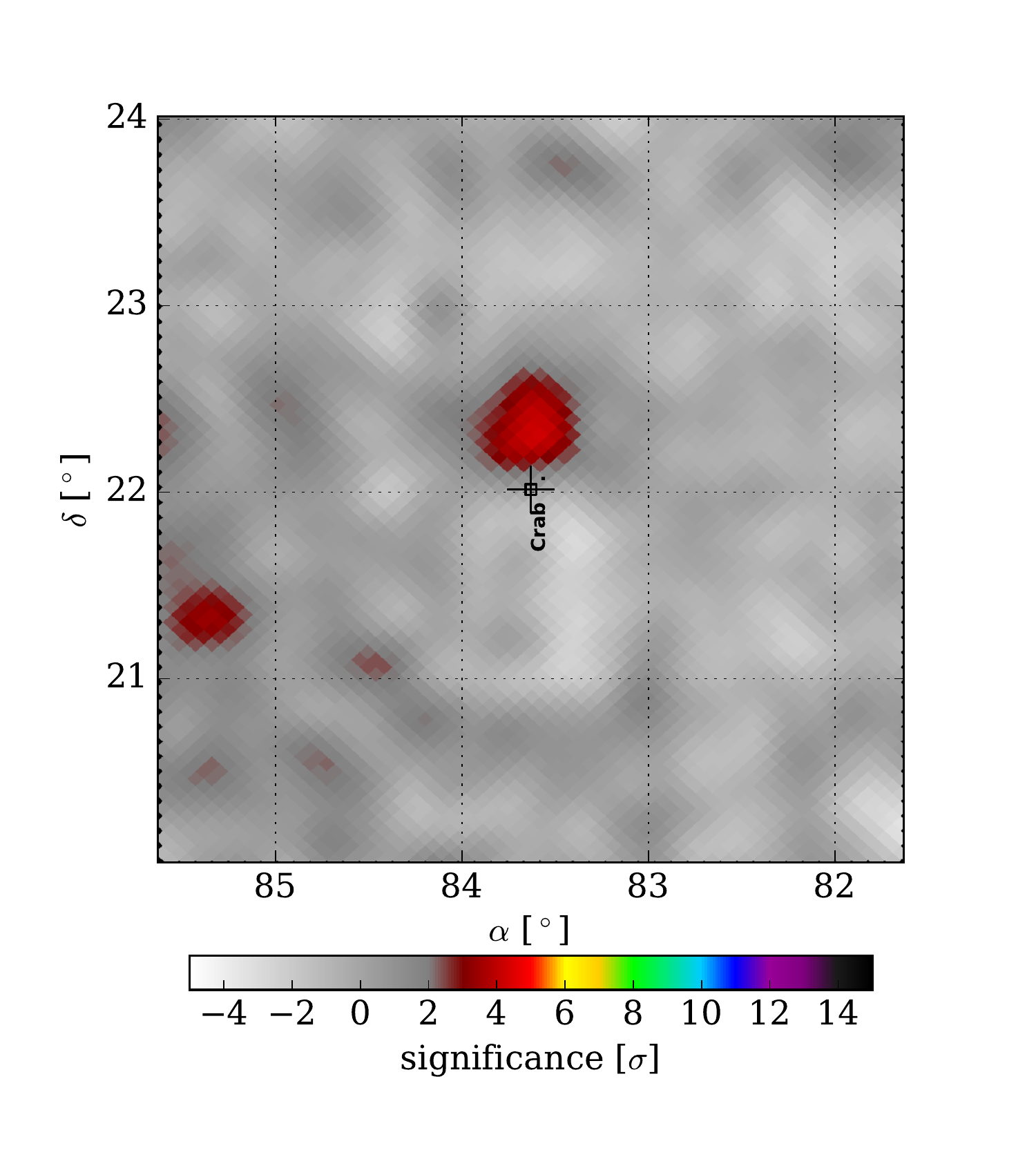}\put(15,83){\color{magenta}{\textbf{PRELIMINARY}}}\end{overpic}\label{fig:d01}}
\subfloat[]
	[Source offset by \ang{0.1}.]
	{\begin{overpic}[width=0.36\textwidth]{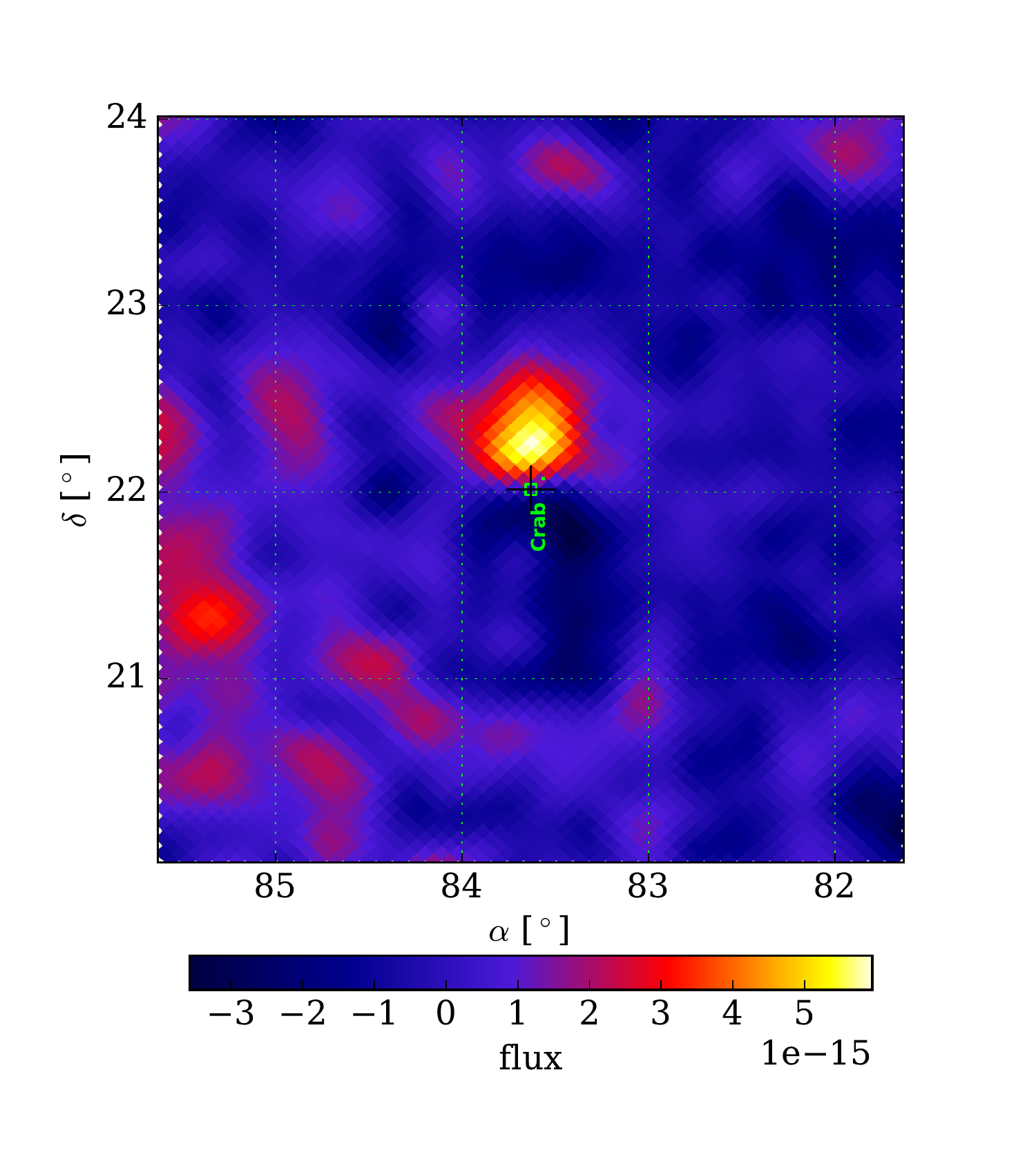}\put(15,83){\color{magenta}{\textbf{PRELIMINARY}}}\end{overpic}\label{fig:d01Flux}}

\subfloat[]
	[Source offset by \ang{0.1} and PSF scaled by 1.1.]
	{\begin{overpic}[width=0.36\textwidth]{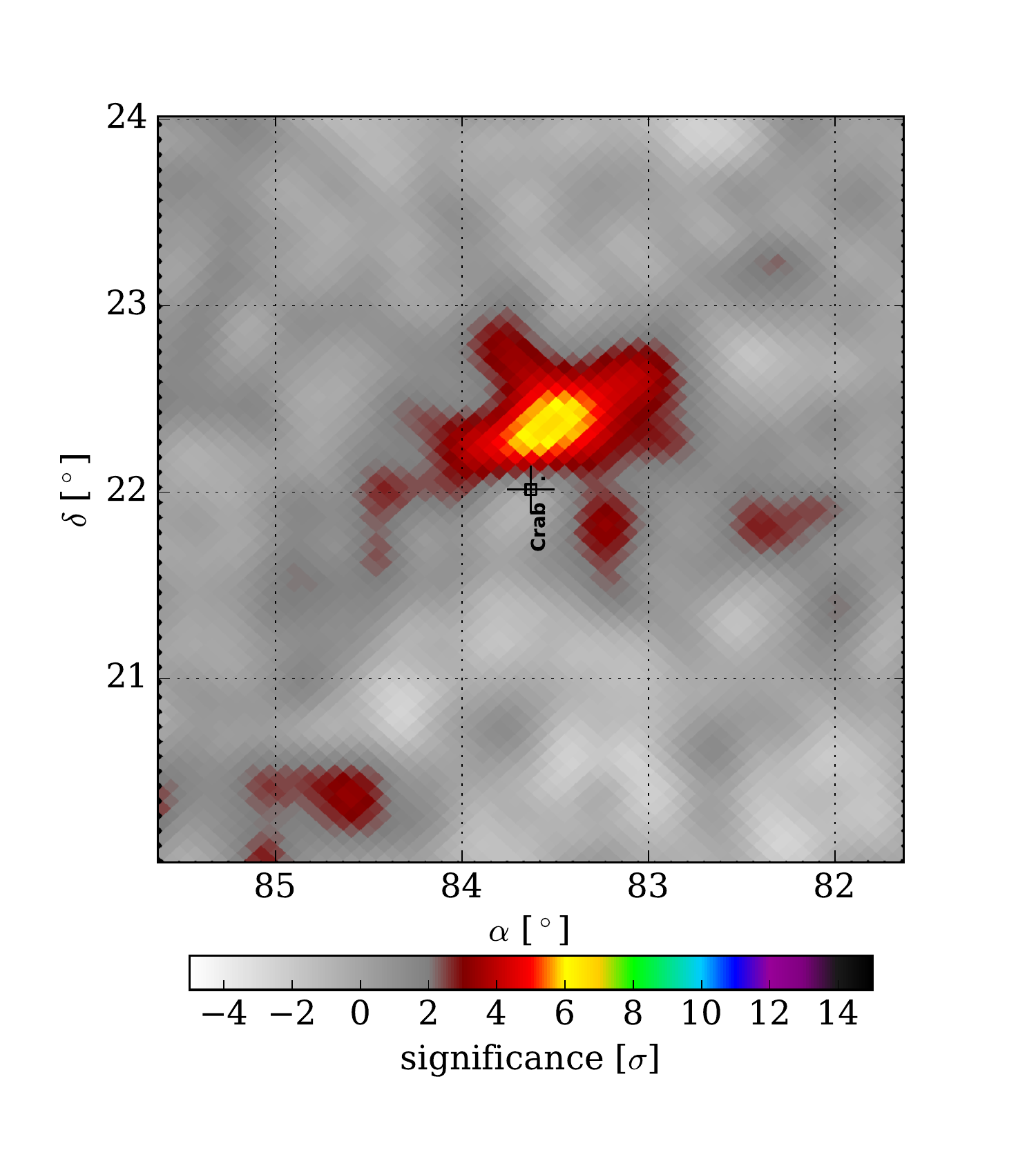}\put(15,83){\color{magenta}{\textbf{PRELIMINARY}}}\end{overpic}\label{fig:d01f11}}
\subfloat[]
	[Source elongated by \ang{0.1}.]
	{\begin{overpic}[width=0.36\textwidth]{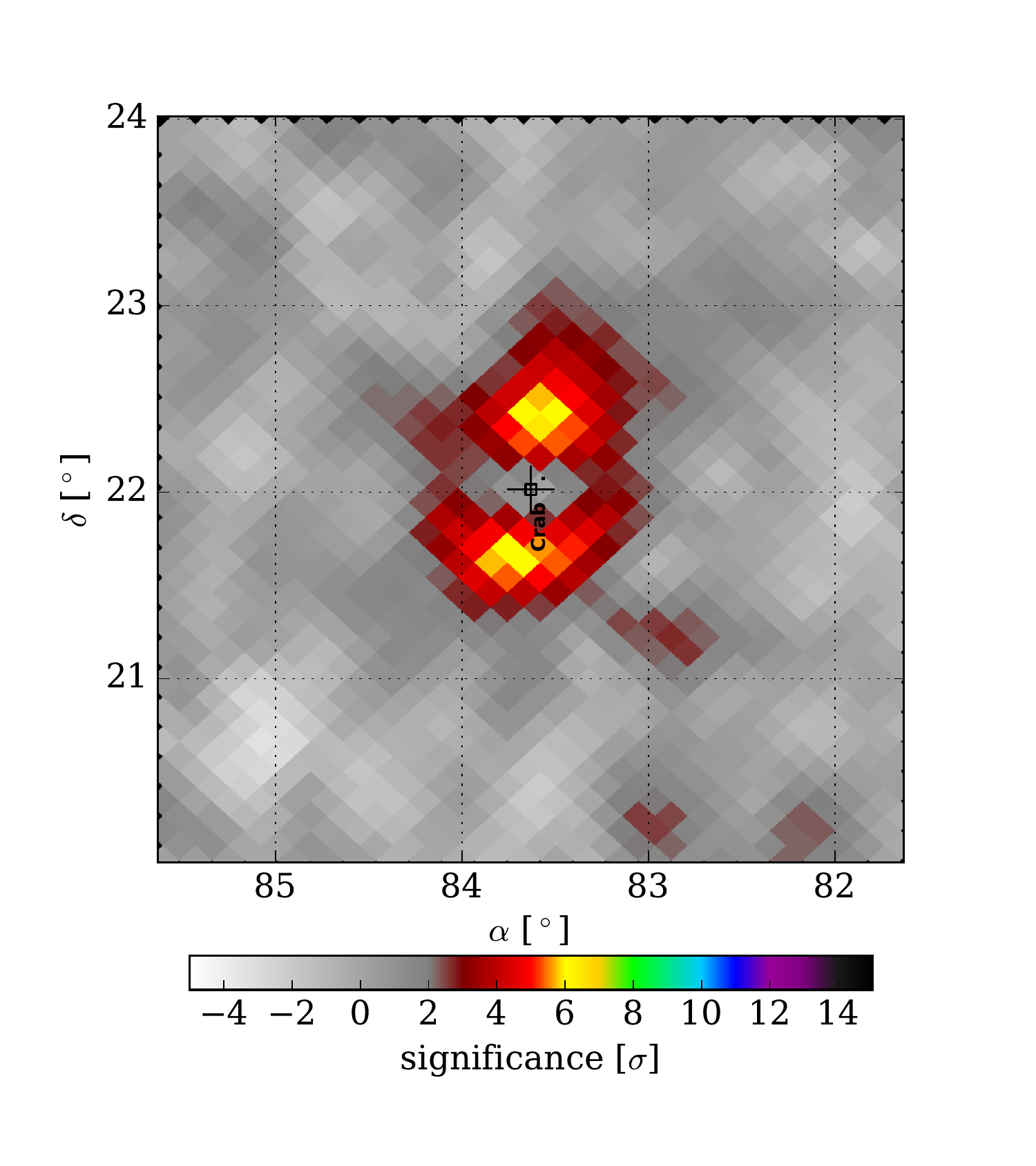}\put(15,83){\color{magenta}{\textbf{PRELIMINARY}}}\end{overpic}\label{fig:e01}}

%\subfloat[]
%	[Source elongated \ang{0.2}.]
%	{\includegraphics[width=0.36\textwidth]{{../../scripts/3ML/fluxes/mapResiduals_PS_fix_FakeEllipse_dRA0_dDec0.0_EllA0.1_Ellb0.12_107_Sig}.pdf}\label{fig:e02}}
\caption{Significance and flux maps of the residuals after subtracting the best-fit point source model from the injected source. The flux values correspond to the differential flux at \SI{7}{\TeV}, for a power-law spectum with index \num{-2.63}.}
\label{fig:maps}
\end{figure}

For each injected source map, a point source model was fit to the injected source and the predicted counts from the best-fit source model were subtracted. Significance maps for these residuals were produced according to the procedure outlined in \cref{sec:analysis}. Some example skymaps can be seen in \cref{fig:maps}.

As expected, if the PSF used for injecting the source is narrower than the one used for the analysis, a ring-like deficit appears around the source position (c.f. \cref{fig:f09}). In contrast, if the injected PSF was wider than the one used for the analysis, a ring-like excess structure appears around the source (c.f. \cref{fig:f11}). Small shifts in the source offset can cause a significant excess, which appears like a second point source (c.f. \cref{fig:d01}). The flux from this `extra' source corresponds to \SI{2}{\percent} of the injected flux (see \cref{fig:d01Flux}).

A combination of a small source offset and a mis-modeled PSF can cause an even more significant excess to appear (c.f. \cref{fig:d01f11}).

We also tested effects of a possibly asymmetric PSF. For this test, the PSF was reduced by \ang{0.1} in each $f_{hit}$ bin and a 2-dimensional Gaussian source with width \ang{0.1} and length \ang{0.1} and \ang{0.2} respectively was injected. One or two significant excesses may appear (c.f. \cref{fig:e01}).

In many example skymaps, the excesses cross the $5\sigma$ significance criterium usually used to claim discovery of a new source. The flux normalizations associated with these excesses correspond to a small percentage (up to \SI{5}{\percent}) of the injected source flux.

\section{Conclusions}
With the HAWC instrument providing unprecented coverage of the TeV gamma-ray sky, source modeling and source subtraction are important tools for the analysis, for example to disentangle complex regions such as the Cygnus region. Measured point source fluxes in the 2HWC catalog may have contaminations of up to \SI{30}{\percent} from Galactic diffuse emission \cite{2HWC}, and the study of diffuse emission with HAWC is currently under way. 

All these studies rely on modeling the response of the HAWC detector to gamma-ray emission. In particular, knowledge of the point spread function is crucial. We have shown here that a systematic error of \SI{10}{\percent} in the angular resolution has a measurable effect on the fitted spectral parameter for strong sources like the Crab Nebula, but these effects are accounted for in the systematic uncertainties.

Additionally, inadequate modeling of the PSF can cause excesses or deficits to appear in the excess maps after subtracting the best-fit source model predictions. For the current unceratinties in the angular resolution and pointing offset, these excess correspond to up to \SI{5}{\percent} of the subtracted source. For strong sources like the Crab Nebula, event this relatively small fraction can lead to significant ($>5\sigma$) excesses. 

Thus, all studies that rely on source subtraction or fitting multiple sources close together should take these effects into account. Efforts to improve the angular resolution and the modeling of the PSF are currently under way, which will improve the performance of the HAWC observatory, especially regarding studies of diffuse emission or of regions with overlapping sources.

\section*{Acknowledgments}
We acknowledge the support from: the US National Science Foundation (NSF); the US Department of Energy Office of High-Energy Physics; the Laboratory Directed Research and Development (LDRD) program of Los Alamos National Laboratory; Consejo Nacional de Ciencia y Tecnolog\'{\i}a (CONACyT), Mexico (grants 271051, 232656, 167281, 260378, 179588, 239762, 254964, 271737, 258865, 243290); Red HAWC, Mexico; DGAPA-UNAM (grants RG100414, IN111315, IN111716-3, IA102715, 109916); VIEP-BUAP; the University of Wisconsin Alumni Research Foundation; the Institute of Geophysics, Planetary Physics, and Signatures at Los Alamos National Laboratory; Polish Science Centre grant DEC-2014/13/B/ST9/945.

\setlength{\bibsep}{0pt plus 0.3ex}
\bibliographystyle{JHEP}
\bibliography{hawc.bib}

\end{document}